\documentclass[preprint,prd,amssymb,floatfix]{revtex4}

\usepackage{graphicx}

\begin{document}

\title{Quantum Stress Tensor Fluctuation Effects in Inflationary Cosmology}

\author{Jen-Tsung Hsiang}\email{cosmology@gmail.com}
\affiliation{Department of Physics, National Dong Hwa University,
Hualien 94701, Taiwan}

\author{Chun-Hsien Wu}
\email{chunwu@scu.edu.tw}
\affiliation{Department of Physics, Soochow University \\
70 Linhsi Road, Shihlin, Taipei 111, Taiwan}

\author{L.H. Ford}
\email{ford@cosmos.phy.tufts.edu}
\affiliation{Institute of Cosmology,
Department of Physics and Astronomy, \\
Tufts University, Medford, MA 02155, USA}

\author{Kin-Wang Ng}
\email{nkw@phys.sinica.edu.tw}
\affiliation{Institute of Physics,
Academia Sinica, Nankang, Taipei 11529, Taiwan}

\begin{abstract}
We review several related investigations of the effects of the quantum stress tensor
of a conformal field in inflationary cosmology. Particular attention will be paid to the
effects of quantum stress tensor fluctuations as a source of density and tensor perturbations in inflationary models. These effects can possibly depend upon the total expansion factor during inflation, and hence be much larger than one might otherwise
expect. They have the potential to contribute a non-scale invariant and non-Gaussian
component to the primordial spectrum of perturbations, and might be observable. 
\end{abstract}

\maketitle

\baselineskip=14pt 

\section{Introduction}

The gravitational effects of quantum matter fields are often described by a
semiclassical theory whereby a suitable renormalized expectation value
of the quantum stress tensor operator becomes the source of gravity. The
semiclassical Einstein equation may be written as
\begin{equation}
G_{\mu\nu} = 8 \pi \ell_p^2 \langle T_{\mu\nu} \rangle \,,
\end{equation}
where $\ell_p$ is the Planck length and $\langle T_{\mu\nu} \rangle$
is the renormalized expectation value. This theory clearly has a wide
domain of applicability, as it includes classical general relativity theory as a 
special case. Furthermore, it is expected to be useful in describing 
non-classical effects, such as quantum violation of the classical energy conditions.

However, the semiclassical theory is limited in that it does not describe
quantum fluctuations of gravity. These fluctuations can arise directly from
the dynamical degrees of freedom of the gravitational field itself, the ``active"
fluctuations. They can also be driven by quantum fluctuations of the quantum
stress tensor, the ``passive" fluctuations. The latter will be the primary focus
of this paper.

The fluctuations of
quantum stress tensors and their physical effects have been discussed
by several authors in recent 
years~\cite{WF01,Borgman,Stochastic,FW04,FR05,TF06,PRV2009}. For a recent review
with further references, see Ref.~\cite{FW07}. Quantum stress tensor 
fluctuations necessarily have a skewed, highly non-Gaussian,
probability distribution, although the explicit form of this
distribution has only been found in two-dimensional spacetime 
models~\cite{FFR10}. The basic features of the probability distribution are a
lower bound and an infinite, positive tail.  The lower bound is at the quantum 
inequality bound for expectation values in an arbitrary state.
For the present purposes, the most important
result is the non-Gaussian character, which implies that cosmological
perturbations driven by stress tensor fluctuations will be non-Gaussian. 
Here we will summarize several related pieces of work and try to illustrate
their interconnections, but for more details, the reader is referred to the original
articles.

\section{The Spacetime Geometry of Inflationary Cosmology}

Here we briefly sketch the assumptions about the spacetime in the version
of inflation which we consider. We take the metric to be that of a spatially flat
Friedmann-Robertson-Walker model
\begin{equation}
ds^2 = -dt^2 +a^2(t)\,  (dx^2 +dy^2 +dz^2)
= a^2(\eta)\,(-d\eta^2 + dx^2 +dy^2 +dz^2)\,.  \label{eq:metric}
\end{equation}
Here $t$ is the comoving time, and $\eta$ the conformal time.
The inflationary period will be taken to be de Sitter spacetime with approximately
constant curvature, so that
\begin{equation}
a= e^{H t} = -\frac{1}{H \eta} \,.
\end{equation}
We may set $a=1$ at the end of inflation, which occurs at $t=0$, or $\eta =-1/H$.
Most inflationary models are insensitive to initial conditions, but that will not be
the case for the effects which we discuss. It will be necessary to impose an
initial condition that the effects of quantum fluctuations vanish at some initial
time, which will be taken to be $\eta = \eta_0$. The total expansion factor
of the universe from this time to the end of inflation is
\begin{equation}
S = \frac{1}{H |\eta_0|} \,.
\end{equation}
This factor will play a key role in our subsequent discussion. 

We assume that inflation ends quickly, with efficient reheating to a subsequent
radiation dominated epoch. This assumption is not crucial, but simplifies the
discussion. In this case, the radiation immediately after inflation is described
by a reheating energy of $E_R$, where
 \begin{equation}
H^2 = \frac{8 \pi}{3}\, \ell_p^2 \, E_R^4\,.
\end{equation}

\section{Semiclassical Effects on Gravity Waves} 

In this section, we will discuss an effect in the semiclassical theory, dealing
with the expectation value of a stress tensor rather than its fluctuations.
However the result has features in common with the stress tensor
fluctuation effects to be discussed later. The effects of $\langle T_{\mu\nu} \rangle$
for a conformal field on the propagation of gravity waves in de Sitter spacetime
was recently treated in Ref.~\cite{HFLY10}. The calculations were based on a formalism developed by Horowitz and Wald~\cite{HW}. It is assumed that at early times,
$\eta < \eta_0$, there is a linearly polarized plane gravitational wave of the form
\begin{equation}
h_\mu ^\nu = c_0\, e_\mu ^\nu \,(1+i k \eta)\,
{\rm e}^{i(\mathbf{k} \cdot \mathbf{x} - k \eta)}\,, 
  \label{eq:grav_wave}
\end{equation}
where $c_0$ is a constant, ${\rm e}_\mu ^\nu$ is the polarization tensor, and
$k$ is the wavenumber.  Next the coupling to the expectation value of the quantized
matter field is assumed to be switched on at $\eta=\eta_0$. 
Its effect is to add a correction term
${h'}_\mu^\nu(x)$, which has the same functional form as  ${h}_\mu^\nu(x)$,
but differs in amplitude and phase. Most importantly, its amplitude at the end of
inflation depends upon the expansion factor $S$ and upon $k$.
In the case that the
conformal field is the electromagnetic field, the fractional correction is
\begin{equation}
\Gamma = \left|\frac{{h'}_\mu ^{~\nu}}{h_\mu ^{~\nu}}\right| =
\frac{1}{5\pi} \ell_p^2 \, H\,S\,k \,.
\label{eq:hprime}
\end{equation}
This effect grows with increasing
$S$ and $k$, but its total magnitude is limited by the requirement
that $\Gamma \alt 1$ for the one-loop approximation to hold.

Nonetheless, this effect could have observational consequences in the form
of a modification of the tensor perturbations predicted by inflation.
Inflationary models predict a nearly scale invariant  spectrum of both
scalar and tensor perturbations, both of which arise from the quantum
fluctuations of nearly free fields and are Gaussian in character. 
The tensor perturbations
result from the fluctuations of quantized linear perturbations of de~Sitter
spacetime~\cite{Starobinsky79,AW84,Allen88}. They have not yet been observed,
but are expected to leave signatures in the cosmic microwave background
that could be seen in the future. The effect described by Eq.~(\ref{eq:hprime})
modifies the normalization of the vacuum graviton modes in a way which
breaks the scale invariance, and increases the power on shorter wavelengths.

\section{Negative Power Spectra}

One of the remarkable properties of quantum stress tensor fluctuations is that
they can introduce negative power spectra of fluctuations. In most statistical
processes, this is not possible.  
The well-known  Wiener-Khinchin~\cite{Wiener,Khinchin} theorem 
states that the Fourier
transform of a correlation function is a power spectrum. 
A corollary of this theorem is that the power spectrum can normally
be written as the expectation value of a squared quantity, and
hence must be positive. However, the latter result can fail in 
quantum field theory, and negative power spectra are possible~\cite{HWF10}. 
The basic reason is that quantum correlation functions can be highly
singular at coincident points, and the expectation value in the 
Wiener-Khinchin theorem may not exist. 

A simple example is the spectrum of vacuum energy density fluctuations of the 
quantized electromagnetic field in flat spacetime. The energy density correlation 
function is
\begin{equation}
C_{0}(\tau,r)= 
\langle \rho(t,{\bf x}) \,\rho(t',{\bf x'})\rangle =
\frac{(\tau^{2}+3r^{2})(r^{2}+3\tau^{2})}
{\pi^{4}(r^{2}- \tau^{2})^{6}}\,,
\label{eq:em_corr}
\end{equation}
where $\tau = t-t'$ and $r = |{\bf x} - {\bf x'}|$. 
Its spatial Fourier transform  is
\begin{equation}
\hat{C}_{0}(\tau,k) = 
-\frac{k^4 \, \sin(k\, \tau)}{960 \pi^5\, \tau} \,,
\label{eq:em_corr_k}
\end{equation}
and the power spectrum is
\begin{equation}
P_{0}(k) = \hat{C}_{0}(0,k) =  -\frac{k^5}{960 \pi^5}\,.
\label{eq:EM-power}
\end{equation}
The negative sign in the power spectrum essentially interchanges
correlations and anticorrelations, compared to what one would have with
a positive spectrum of the same functional form.

\section{Density Perturbations in Inflation}

Just as fluctuations of the free graviton field in de Sitter spacetime can
lead to observable tensor perturbations, fluctuations of the inflaton field
can produce density 
perturbations~\cite{MC81,GP82,Hawking82,Starobinsky82,BST83}.
The resulting spectrum of nearly scale invariant, Gaussian fluctuations
has apparently been observed in the temperature fluctuations
of the cosmic microwave background~\cite{WMAP}.  However, quantum
stress tensor fluctuations can produce an additional, non-scale invariant,
non-Gaussian contribution, which was studied in Refs.~\cite{WNF07,FMNWW10}. 
This effect arises from the quantum fluctuations of the comoving energy density of a conformal field in its vacuum state. In a dynamic model treated in 
Ref.~\cite{FMNWW10}, these fluctuations couple to the dynamics of the inflaton
field and produce a spectrum of density perturbations proportional to $S^2$,
\begin{equation}
{\cal P}(k) = \frac{8 \ell_p^4 \, k^2\, S^2}{75} \,.
\end{equation}
Note that
\begin{equation}
 {\cal P}(k) = 4 \pi k^3 \, P(k) \,,
\end{equation}
is the quantity usually called the power spectrum in cosmology, and which
is independent of $k$ for a scale invariant spectrum. Thus the effect of the
quantum stress tensor fluctuations is a blue-tilted, non-Gaussian contribution.

The fact that this contribution has not yet been observed can be interpreted
as placing an upper bound on $S$, which is found to be
\begin{equation}
S \alt 10^{42} \left(\frac{10^{12}~{\rm GeV}}{E_R}\right)^3 \, .
\label{eq:bound}
\end{equation}
This bound is consistent with adequate inflation to solve the horizon and flatness
problems, which requires $S > 10^{23}$.

\section{Gravity Waves from Stress Tensor Fluctuations in Inflation}

In addition to the scalar (density) perturbations, stress tensor fluctuations can also
create tensor perturbations, which are gravity waves~\cite{WHFN}. Here the fluctuations
of spatial components of the stress tensor can couple to the transverse, tracefree
components of the gravitational field. The result at the end of inflation is a
spectrum of gravity wave fluctuations with a power spectrum given by
\begin{equation}
{ \cal P}(k) = -\frac{4 \ell_p^4 \,k^2\,H^2\, S^2}{3 \pi} (1 + k^2\, H^{-2})
\,.
\label{eq:power-sudden2}
\end{equation}  
This is an example of a negative power spectrum, as well as one which is
strongly blue-tilted. This result assumes that the coupling between the
quantum stress tensor and the gravitational field is switched on suddenly at
$\eta=\eta_0$. However, more gradual switching leads to qualitatively
similar results.  If $S$ were sufficiently large, then the resulting tensor perturbations
in the CMB spectrum should have already been observed. This leads to the
constraint
\begin{equation}
 S \alt  10^{46} \,  \left(\frac{10^{12} GeV}{E_R}\right)^3\,,
 \end{equation}
which is somewhat weaker than the bound, Eq.~(\ref{eq:bound}), which comes
from density perturbations.

However, gravity wave fluctuations are potentially detectable at much smaller
scales than are the density perturbations. The shorter wavelengths of the primordial
density perturbation spectrum have presumably been erased by nonlinear classical
effects by now,  but gravity waves interact very weakly and should still exist.
This raises the possibility of  detecting the primordial gravity waves from stress
tensor fluctuations in gravity wave detectors, in the form of background noise
corresponding to the spectrum given in Eq.~(\ref{eq:power-sudden2}).
LIGO has placed limits~\cite{LIGO}
 of $h \alt 10^{-24}$ on scales of the order of $10^2\,km$, leading to the constraint
\begin{equation}
S < 10^{23} \, \left(\frac{10^{10} GeV}{E_R}\right)^3 \,.
\end{equation}
This result is  compatible with adequate inflation to solve the horizon
and flatness problems only if
\begin{equation}
E_R \alt 10^{10}\, GeV \,.
\label{eq:ER_constraint}
\end{equation}
This would  put a non-trivial constraint on inflationary models. An even more exciting possibility is that Earth or space-based gravity wave detectors might have a
chance of detecting the effects of quantum stress tensor fluctuations during 
inflation.

\section{The Transplanckian Issue}

There is an important qualification to all of the results which depend upon $S$,
the expansion factor during inflation. This is that quantum field modes above the
Planck scale in energy are required. Consider, for example, the example in the
previous section of a gravity wave with a present proper wavelength of $100 km$.
If we assume $E_R \approx  10^{10}\, GeV$, then there has been an expansion
by a factor of about $10^{23}$ since reheating to redshift to the current CMB
temperature. This,
combined with an additional expansion of at least $10^{23}$ during inflation,
implies that this mode would have had a proper wavelength of about 
$10^{-3} \ell_p$ at the beginning of inflation. The quantum field would have to
have fluctuations on this scale to create such a gravity wave mode.

This raises questions as to whether the framework of quantum field theory on
a fixed or weakly fluctuating background can be trusted at this scale. However,
an even more extreme version of the transplanckian issue arises in black hole
thermodynamics. Hawking's~\cite{Hawking} derivation of  of black hole radiance
relies 
upon modes which begin far above the Planck energy. The fact that the Hawking effect gives a
beautiful unification of gravity, thermodynamics, and quantum theory can be 
considered to be a powerful argument to take  transplanckian modes seriously.
It is true that it is possible to derive the Hawking effect without  transplanckian 
modes~\cite{Unruh,CJ96}, but only at the price of introducing modified
dispersion relations which break local Lorentz symmetry and hence postulate
new physics. An analogous prescription in cosmology is to consider modes only
after they redshift below the Planck energy in the comoving 
frame~\cite{FMNWW10}. There has been an extensive discussion of the possible role of
transplanckian modes in inflationary cosmology. (See Ref.~\cite{FMNWW10}
for a lengthy list of references.) 

The dependence of the gravity wave spectrum upon a positive power of $S$ might
seems to contradict a theorem due to Weinberg~\cite{Weinberg}, which was generalized
by  Chaicherdsakul~\cite{Chai07}. This theorem states that radiative corrections during
inflation should not grow faster than a logarithm of the scale factor. However, as
was discussed in more detail in  Ref.~\cite{FMNWW10}, the effect of
 density perturbations growing as a power of $S$ is really due to
high frequency modes at the initial time, and is hence always large rather than growing.
This applies to  all the effects discussed in the present paper.  

\section{Discussion}

Here we have reviewed three distinct, but related, quantum effects in inflation which
depend upon the expansion factor $S$ during inflation. The first is the effect
of the expectation value of a conformal field upon the amplitude of gravity
waves in de Sitter spacetime~\cite{HFLY10}. The second is the effect of energy
density fluctuations of such a field upon the spectrum of primordial density
perturbations~\cite{WNF07,FMNWW10}. The third effect is the role of quantum
stress tensor fluctuations in generating a spectrum of primordial gravity 
waves~\cite{WHFN}.
The latter two effects are associated with  non-Gaussian fluctuations, and
all three effects grow as $S$ increases  and produce  a blue-tilted spectrum,
in which there is more power on shorter wavelengths. The gravity waves
produced by stress tensor fluctuations are an example of a negative power
spectrum, which is usually forbidden, but is possible in quantum field 
theory~\cite{HWF10}. 

All three effects have the potential to produce observable effects, either in the
cosmic microwave background radiation, or in gravity wave detectors. However,
all three effects also rely upon the existence of transplanckian modes. As such,
they have the possibility to open an observational window on transplanckian physics.

\begin{acknowledgments}
We would like to thank Da-Shin Lee and  Hoi-Lai Yu
 for valuable discussions. 
This work is partially supported by the National Science Council, Taiwan, ROC 
 under the Grants NSC98-2112-M-001-009-MY3 and NSC99-2112-M-031-002-MY3,
  and by the U.S. National Science Foundation under Grant PHY-0855360.  
\end{acknowledgments}


\begin{thebibliography}{}


\bibitem{WF01} C.-H. Wu and L.H. Ford, Phys. Rev. D {\bf 64}, 045010 (2001),
quant-ph/0012144.

\bibitem{Borgman} J. Borgman and L.H. Ford, Phys. Rev. D {\bf 70}
 064032 (2004), gr-qc/0307043.

\bibitem{Stochastic} B.L. Hu and E. Verdaguer, Living Rev. Rel. {\bf 7},
3 (2004), gr-qc/0307032.

\bibitem{FW04} L.H. Ford and R.P. Woodard, Class. Quant. Grav. {\bf 22}, 
1637 (2005), gr-qc/0411003.

\bibitem{FR05} L.H. Ford and T.A. Roman, Phys. Rev. D {\bf 72}, 
105010  (2005),  gr-qc/0506026

\bibitem{TF06} R.T. Thompson and L.H. Ford,  Phys. Rev. D {\bf 74},
024012 (2006), gr-qc/0601137.

\bibitem{PRV2009} G. Perez-Nadal, A. Roura   and E.
Verdaguer,  JCAP 1005:036 (2010), arXiv:0911.4870.

\bibitem{FW07} L.H. Ford and C.H. Wu, AIP Conf. Proc. {\bf 977} 145
  (2008), arXiv:0710.3787.

\bibitem{FFR10}  C.J. Fewster, L.H. Ford, and T.A. Roman, Phys. Rev. D {\bf 81},
121901(R) (2010),  arXiv:1004.0179.

\bibitem{HFLY10} J.-T. Hsiang, L.H. Ford, D.-S. Lee, and  H.-L. Yu, Phys. Rev. D,
{\bf 83}, 084027 (2011),  arXiv:1012.1582.

\bibitem{HW}  G.T. Horowitz and R.M. Wald,  Phys. Rev. D {\bf 21}, 1462 (1980);
 {\bf 25}, 3408 (1982).

\bibitem{Starobinsky79} A.A. Starobinsky, JETP Lett. {\bf 30}, 682 (1979).

\bibitem{AW84} L.F. Abbott and M.B. Wise, Nucl. Phys. B {\bf 244}, 541 (1984).

\bibitem{Allen88} B. Allen, Phys. Rev. D {\bf 37}, 2078 (1988).

\bibitem{Wiener} N. Wiener, Acta. Math, Stockholm {\bf 55}, 117 (1930).

\bibitem{Khinchin} A. Khinchin, Math. Ann. {\bf 109}, 604 (1934).

\bibitem{HWF10} J.-T. Hsiang, C.-H. Wu, and L. H. Ford, Phys. Lett. A {\bf 375},
2296 (2011),  arXiv:1012.3226.

\bibitem{MC81} V. Mukhanov and G. Chibisov, JETP Lett. {\bf 33}, 532 (1981).

\bibitem{GP82} A.H. Guth and S.-Y. Pi, Phys. Rev. Lett. {\bf 49}, 1110 (1982).

\bibitem{Hawking82} S.W. Hawking, Phys. Lett. B {\bf 115}, 295 (1982).

\bibitem{Starobinsky82}  A.A. Starobinsky,
Phys. Lett. B {\bf 117}, 175 (1982).

\bibitem{BST83}  J.M. Bardeen, P.J. Steinhardt, and
M.S. Turner, Phys. Rev. D {\bf 28}, 679 (1983).

\bibitem{WMAP} E. Komatsu {\it et al.}, Astrophys. J. Suppl, {\bf
192}, 18 (2011),  arXiv:1001.4538.

\bibitem{WNF07} C.-H. Wu, K.-W. Ng, and L.H. Ford,  Phys. Rev. D 
{\bf 75}, 103502 (2007), gr-qc/0608002. 

\bibitem{FMNWW10} L.H. Ford, S.-P. Miao, K.-W. Ng, R. Woodard, and 
C.-H. Wu,  Phys. Rev. D {\bf 82}, 043501 (2010), arXiv:1005.4530.

\bibitem{WHFN}  C.-H. Wu, J.-T. Hsiang,  L.H. Ford, and K.-W. Ng, 
arXiv:1105.1155.

\bibitem{LIGO} P.J. Sutton, J. Phys. Conf. Ser. {\bf 110}, 062024 (2008).

\bibitem{Hawking} S.W. Hawking, Commun. Math. Phys. {\bf 43}, 199 (1975).

\bibitem{Unruh} W.G. Unruh,  Phys. Rev. D {\bf 51}, 2827 (1995), gr-qc/9409008.

\bibitem{CJ96} S. Corley and  T. Jacobson,  Phys. Rev. D {\bf 54}, 1568 (1996),
hep-th/9601073.

\bibitem{Weinberg} S. Weinberg, Phys. Rev. D {\bf 72}, 043514 (2005);
{\bf 74}, 023508 (2006).

\bibitem{Chai07} K. Chaicherdsakul, Phys. Rev. D {\bf 75}, 063522 (2007).




\end{thebibliography}
\end{document}